\documentclass[a4paper]{article} 
\usepackage{epsfig}
\usepackage{amsmath}
\usepackage[left=2cm,top=2cm,bottom=3cm]{geometry}

\makeatletter 
\@addtoreset{equation}{section} 
\makeatother 
 
\newcommand{\be}{\begin{equation}} 
\newcommand{\ee}{\end{equation}} 
\newcommand{\bea}{\begin{eqnarray}}
 \newcommand{\eea}{\end{eqnarray}}
 \def\non{\nonumber }
\newcommand{\ov } {\over } 
\newcommand{\p }{\partial }
 \newcommand{\s }{\sigma }

\def\lb{\lambda}

\def\a{\alpha }

\def\lb{\lambda}

 \def\vareps{\varepsilon }
 \def\eps{\epsilon }

\def\Tr{ {\rm Tr}}

\def\appendix#1{   \addtocounter{section}{1}   \setcounter{equation}{0}   
\renewcommand{\thesection}{\Alph{section}}   \section*{Appendix \thesection\protect\indent \parbox[t]{11.15cm}   {#1} }   \addcontentsline{toc}{section}{Appendix \thesection\ \ \ #1}   } 

\def\appendix#1{
  \addtocounter{section}{1}
  \setcounter{equation}{0}
  \renewcommand{\thesection}{\Alph{section}}
  \section*{Appendix \thesection\protect\indent \parbox[t]{11.15cm}
  {#1} }
  \addcontentsline{toc}{section}{Appendix \thesection\ \ \ #1}
  }

\begin{document}

\null\vskip-24pt 
      \vskip-1pt
\vskip 1truecm
\begin{center}
\vskip 0.2truecm {\Large\bf 
Non-linear theory for multiple M2 branes\\
 }
\vskip 0.2truecm

\vskip 0.7truecm
\vskip 0.7truecm

{\bf Roberto Iengo$^a$ and Jorge G. Russo$^{b,c}$}\\
\vskip 0.4truecm
\vskip 0.4truecm

${}^a${\it  International School for Advanced Studies (SISSA)\\
Via Beirut 2-4, I-34013 Trieste, Italy} \\
{\it  INFN, Sezione di Trieste}

\medskip

$^{b}${\it 
Instituci\' o Catalana de Recerca i Estudis Avan\c{c}ats (ICREA)

  \medskip

$^{c}${\it 
Departament ECM and Institut de Ciencies del Cosmos,\\
Facultat de F\'\i sica, Universitat de Barcelona,\\
Diagonal 647, 08028 Barcelona, Spain} 
} 
 
\end{center}
\vskip 0.2truecm 

\noindent\centerline{\bf Abstract}

We present a manifestly $SO(8)$ invariant non-linear Lagrangian for describing
the non-abelian dynamics of the bosonic degrees of freedom of $N$ coinciding M2 branes in flat spacetime.
The theory exhibits a gauge symmetry structure of the $BF$ type (semidirect product of $SU(N)$ and translations) and at 
low energies it reduces exactly to the bosonic part of the Lorentzian Bagger-Lambert Lagrangian for group $SU(N)$. 
There are eight scalar fields satisfying a free-scalar equation. When one of them takes a large expectation value,
the non-linear Lagrangian gets simplified and the theory can be connected to the non-abelian
Lagrangian describing the dynamics of $N$ coinciding D2 branes.
As an application, we show that the BPS fuzzy funnel solution describing M2 branes ending into a single M5 brane
is an exact solution of the non-linear system.


\newpage

\section{  Introduction}

\setcounter{equation}{0}

Understanding the dynamics of multiple M2 branes may reveal important aspects
of the  microscopic structure of M-theory.
Recently several models for M2 brane dynamics with explicit Lagrangian description have appeared in the literature.
In \cite{BL1} 
Bagger and Lambert found a Lagrangian with maximal superconformal invariance containing the expected
degrees of freedom of M2 branes (see also \cite{Basu,gus}).
The construction uses an algebraic structure called Lie 3-algebra, parametrized by structure constants $f^{abc}_{\ \ \ \ d}$, 
and  a bi-invariant metric $h^{ab}$. The structure constants must satisfy a quadratic condition which turns out to be quite restrictive.
It was shown in \cite{papa,gaun} that for a positive definite metric $h^{ab}$ the  known example $f^{abcd}\propto \varepsilon^{abcd}$ is essentially unique,
leading to a model with local $SO(4)$ invariance 
which can be interpreted as describing two M2 branes in an $R^8/Z_2$ orbifold background \cite{Tong,Distler}.

In \cite{GMR,ver1,Mat} it was shown that if the metric $h^{ab}$ has Lorentzian signature, then one can construct superconformal models for any Lie algebra. 
In particular, choosing this Lie algebra to be $su(N)$, one obtains an ${\cal N}=8$ superconformal invariant Lagrangian, proposed to describe the dynamics of $N$
M2 branes in flat spacetime. By giving an expectation value to one of the scalar fields through the procedure found in \cite{mukhi}, one can indeed 
show \cite{GMR,Mat} that in the IR regime (corresponding to a large expectation value) 
the model reduces to the maximally supersymmetric Yang-Mills Lagrangian describing the low-energy dynamics of $N$ D2 branes.
The bosonic part of the Lagrangian is given by
\be
L ={\rm Tr}\bigg( {1\over 2} \eps^{\mu\nu\rho } B_\mu F_{\nu\rho } - {1\over 2} \hat D_\mu X^I \hat D^\mu X^I +{1\over 12} M^{IJK} M^{IJK}\bigg)
+\big( \partial_\mu  X_-^I - {\rm Tr}[B_\mu X^I]\big) \partial^\mu X_+^I \ ,
\label{lag1}
\ee
where the fields $A_\mu =A_\mu^a T^a$,
$B_\mu =B_\mu^a T^a$, $X^I= X^{Ia} T^a$, transform in the adjoint of $su(N)$,  whereas $X_\pm ^I$ are $su(N)$ singlets.
We take hermitian $N\times N$ matrices $T^a$, $a=1,...,N^2-1$, satisfying ${\rm Tr}[T^a T^b]=N\delta_{ab}$. 
We have also introduced the notation
\be
M^{IJK} \equiv X^I_{+}[X^J,X^K]+ X^J_{+}[X^K,X^I]+X^K_{+}[X^I,X^J]\ ,
\label{mijk}
\ee
\be
\hat D_\mu X^I \equiv D_\mu X^I-X^I_{+} B_\mu , \ \  \ \  D_\mu X^I\equiv\p_\mu X^I+i[A_\mu, X^I]\ .
\label{covdev}
\ee

As a consequence of the Lorentzian signature of $h^{ab}$, there is a field $X_0^I\equiv X_+^I + X_-^I $ with the wrong sign in the kinetic term,
which may lead to violation of unitarity.
Different arguments have been given in \cite{GMR,ver1,Mat} (see also  \cite{cecotti,Fig,ASS,ver3}) that the model may  nevertheless be unitary
due to the particular form of the interactions, which  ensure that $X_+$ and $X_-$ can be integrated out by its equations of motion; they also imply that the ghost-like fields do not
run in  loops of Feynman diagrams. 
The role of the $X_+^I, X_-^I$ fields is to provide a special kind of dressing that leads to the conformal invariance of the model.

A different strategy studied in \cite{schwarz,Gomis2} is to gauging the global translational symmetry $X_-^I\to X_-^I+ c^I$ by means of the introduction of a gauge field $C_\mu ^I$
in a new term in the Lagrangian $-C_\mu ^I \p^\mu X_+^I$. The equation of motion of $C^I_\mu $ then freezes out the mode $X_+^I$ to a constant value. The resulting
 model seems to be essentially equivalent the maximally supersymmetric Yang-Mills Lagrangian describing the low energy dynamics of D2 branes, though
 this has not yet been shown in a complete treatment including calculation of observables (see also \cite{mukhi2}).

In addition to the $SU(N)$ gauge symmetry, the above Lagrangian is invariant under the (non-compact) gauge symmetry transformations
associated with the $B_\mu $ gauge field,
\be
\delta X^I=X^I_{+}\Lambda \ , \ \ \  \delta B_\mu=D_\mu\Lambda \ ,\ \ \ \delta X^I_{+}=0 \ ,\ \ \ \delta X^I_{-}= \Tr ( X^I \Lambda)\ .
\label{Bgauge}
\ee
 The symmetry algebra underlying the model is generated by  $J^a$, $P^a$ satisfying the BF algebra
 \be
 [J^a, J^b] = iC^{ab}_{\ \ \ c} J^c\ ,\qquad [P^a, J^b] = iC^{ab}_{\ \ \ c} P^c\ ,\qquad [P^a, P^b] = 0\ .\
 \label{bfalg}
\ee
where $C^{ab}_{\ \ \ c}$ are (real) structure constants of $su(N)$.

The Lagrangian (\ref{lag1}) is a candidate to  describe M2 brane dynamics in the low-energy approximation.
The full M2 brane dynamics is expected to be described by a non-linear theory which at low energies reduces to (\ref{lag1}) 
and in some limit (discussed below) reduces to the non-linear dynamics of $N$ D2 branes.
The  non-linear Lagrangian describing the dynamics of D branes is not fully understood in the non-abelian case. However, 
there is a concrete Lagrangian for the bosonic degrees of freedom
  \cite{tseytlin,myers} which works quite well up to high orders in $\alpha' $.
{}For flat backgrounds, the  non-abelian D2 brane Lagrangian reduces to 
\be
L=-T \ {\rm STr}\ \sqrt{-\det (\eta_{\mu\nu}+\lambda^2 D_\mu\Phi^i Q^{-1}_{ij}D_\nu \Phi^j+\lambda F_{\mu\nu}) \det Q}\ .
\label{myersL}
\ee 
where STr means symmetrized trace \cite{tseytlin} and  
\be
Q^{ij}=\delta^{ij}+i\lambda [\Phi^i,\Phi^j]\ .
\ee
As usual, $\Phi^ i$ represents the transverse displacements, $\Delta x^i=\lb \Phi^i\ ,\ \lb=2\pi l_s^2$.
For further details we refer to \cite{tseytlin,myers}. The tension is
\be
T={1\over (2\pi)^2 l_s^3 g_s}={1\ov\lambda^2 g_{\rm YM}^2}\ ,\qquad g_{\rm YM}^2={g_s\over l_s}\ .
\label{tenz}
\ee

{}For  a single M2 brane, the classical non-linear dynamics is governed by the supermembrane action \cite{BST}. For multiple M2 branes,
the non-linear action analogous to the non-abelian D brane Lagrangian is not known.
The aim of this paper is to find a non-linear $SO(8)$ invariant Lagrangian for the bosonic degrees of freedom of the M2 branes 
that reduces to the non-abelian D2 brane Lagrangian at large $g_{\rm YM}$ coupling and to the BF membrane Lagrangian (\ref{lag1})  at
low energies.

Another proposal for M2 branes in flat spacetime was presented in \cite{abjm}, called ABJM models, in terms of a Lagrangian that realizes
six supersymmetries (see also \cite{klebanov,schwarz2}).
We have not found a natural ansatz for the non-linear generalization of the ABJM models, so we will not discuss them in this paper.
Some studies of non-linear Lagrangians for M2 branes, which do not overlap with this paper, are in \cite{Xie,kluson}.
It would also be interesting to understand the non-linear theory for the Bagger-Lambert construction based on the Nambu-bracket \cite{mat1,mat2,BT1,BT2}.

This paper is organized as follows.
In section 2 we start with the abelian case. Here one can write two alternative proposals, but only one of them  survives in the non-abelian case.
In section 3 we consider the non-abelian case and propose a non-linear M2 brane Lagrangian with the desired symmetry structure,
which turns out to be directly related to  the non-abelian D2 brane lagrangian when one of the scalar fields is set to a constant value.
In section 4 we check that the supersymmetric funnel of eleven dimensions --~representing a fuzzy M2-M5 brane intersection~--  is an exact solution of our proposal,
and that is not modified by the non-linearities, just as it happens in the D1-D3 brane case \cite{constable1}.
 

\section{From D2 branes to M2 branes in the abelian case }
\setcounter{equation}{0}

The connection between the single D2 brane and the single M2 brane  action was derived in \cite{town}.
Here we will review part of this connection, following \cite{town}, and in addition connect with the recently found 
BF membrane (or ``Lorentzian Bagger-Lambert") theory (\ref{lag1}) based
on the Bagger-Lambert construction. We will only consider the part containing the bosonic fields.
 The BI Lagrangian for a D2 brane in the static gauge is given by
\be
L = -T \sqrt{- \det \left( g_{\mu\nu}   + \lambda F_{\mu\nu} \right) }\ ,
\label{uno}
\ee
where
\be
g_{\mu\nu} = \eta_{\mu\nu} + \lambda^2 \p_\mu \Phi^i \p_\nu \Phi^i  \ ,\qquad i=1,...,7\ .
\ee
By introducing a Lagrange multiplier $p$, this can be written as
\be
L= {1\over 2 p} T^2 \det \big( g_{\mu\nu}\big)   -{1\over 2} p\, \Big(1 + {1\over 2}\lambda^2 |F|^2 \big) \ , 
\label{due}
\ee
where we used the identity for $3\times 3$ matrices
\be
\det \big( g_{\mu\nu} + \lambda F_{\mu\nu} \big) = \det \big( g_{\mu\nu}\big) \ \Big(1 + {1\over 2}\lambda^2 |F|^2 \big) \ , 
\qquad |F|^2 = g^{\mu\rho} g^{\nu\sigma} F_{\mu\nu} F_{\rho\sigma}\ ,
\label{tre} 
\ee 
which applies for any antisymmetric $F_{\mu\nu }$.  Introducing an auxiliary field $B_\mu $, we can write the Lagrangian as
\be
L={1\over 2 p} T^2 \det \big(g_{\mu\nu}\big)\ \big(1+ \lambda^2 g_{\rm YM}^4  B_\mu B_\nu g^{\mu\nu }\big) +{1\over 2}\epsilon^{\mu\nu\rho}   B_\mu F_{\nu\rho}   -{1\over 2} p \ .
\label{quattro}
\ee
This is the standard duality  \cite{nicolai} connecting Chern-Simons and Yang-Mills theory in three dimensions.
Solving the equation for $B_\mu $, substituting in (\ref{quattro}) and using 
\be
|F|^2 = {1\over 2} (\det g)^{-1} g_{\mu\mu'} \epsilon^{\mu\nu\rho}  \epsilon ^{\mu'\nu'\rho'} F_{\nu\rho}F_{\nu'\rho'}\ .
\ee
one can verify that the Lagrangian (\ref{uno}) is reproduced. 

Next, using the identity for $3\times 3$ matrices
\be
\det \big(g_{\mu\nu} + K_\mu K_\nu \big) = \det g_{\mu\nu} \ \Big(1 + K_\mu K_\nu g^{\mu\nu }\big) \ , 
\label{cinque} 
\ee 
we get
\be
L={1\over 2 p} T^2 \det \big(g_{\mu\nu} + \lambda^2 g^4_{\rm YM} B_\mu B_\nu \big) + {1\over 2}\epsilon^{\mu\nu\rho}   B_\mu F_{\nu\rho}   -{1\over 2} p\ .
\label{seii}
\ee
Solving the equation of motion for $p$ we find
\be
L= -T \sqrt{ -\det \big(g_{\mu\nu} + \lambda^2 g^4_{\rm YM}  B_\mu B_\nu \big) } + {1\over 2}\epsilon^{\mu\nu\rho}   B_\mu F_{\nu\rho} \ .
\label{sei}
\ee
Now the equation for $A_\mu $ is solved by $B_\mu =\p_\mu \phi $. This introduces the eight-th scalar field $\Phi^8$ in the Lagrangian, $\Phi^8 \equiv  g^2_{\rm YM} \phi   $.
In order to compare with the Lagrangian (\ref{lag1})  (and to have canonically normalized scalar fields), we introduce new variables $X^I$ by
\be
\Phi^I=g_{\rm YM}X^I\ ,\qquad I=1,...,8\ .
\label{nomal}
\ee
Note that $[X^I]=\mu^{1/2}$ carries the standard dimensionality of a bosonic field in $D=2+1$ and that also $[g_{\rm YM}]=\mu^{1/2}$.  
Therefore, we finally get
\be
L= -T \ \sqrt{ -\det \big(\eta_{\mu\nu} +{1\ov T} \p_\mu X^I \p_\nu X^I\big) }  \ ,
\label{sette}
\ee
where we used $ T^{-1} = \lambda^2 g_{\rm YM}^2$ (see eq. (\ref{tenz})).
The eleven-dimensional 
Planck length scale $l_p$ is related to $T$ by $ T^{-1}=(2\pi)^2l_p^3$ (we used $l_p^3= l_s^3 g_s $).
Thus we find the Lagrangian for a membrane in the static gauge with the expected $SO(8)$ symmetry. 

\medskip

Now we will show that the Lagrangian (\ref{sei}) arises from 
either one of the following non-linear generalizations of the 
(abelian) BF membrane Lagrangian:\footnote{Here and in what follows we ignore gauge fixing terms and corresponding ghost contributions. The discussion will be purely classical.}
\bea
L_1 &=& - T \sqrt{ -\det \left(\eta_{\mu\nu} +{1\over T}  \Big(  \hat D_\mu X^I \hat D_\nu X^I 
- 2\big[ \p_{( \mu } X^I_- -  B_{( \mu } X ^I \big] \p_{ \nu )}  X_+^I \Big)\right) } 
\non\\
&+& {1\over 2}\epsilon^{\mu\nu\rho}   B_\mu F_{\nu\rho} \ ,
\label{otto}
\\
L_2 &=& - T \sqrt{ -\det \left(\eta_{\mu\nu} + {1\ov T} \tilde D_\mu X^I \tilde D_\nu X^I\right) }  
+ {1\over 2}\epsilon^{\mu\nu\rho}   B_\mu F_{\nu\rho}  \non\\
&+& (\p_\mu X^I_{-} - X^I B_\mu)\p^\mu X^I_{+}
-   {X_{+}\cdot X\ov X^2_{+}}\hat D_\mu X^I\p^\mu X^I_{+} + {1\over 2}  ({X_{+}\cdot X\ov X^2_{+}})^2\p_\mu X^I_{+}\p^\mu X^I_{+}\ ,
\label{xotto}
\eea
where, as usual, $A_{( \mu }B_{\nu )} \equiv {1\over 2} \big(A_\mu B_\nu + A_\nu B_\mu\big) $, and
\be
\tilde D_\mu X^I=\hat D_\mu X^I-{X_{+}\cdot X \over X^2_{+}}\p_\mu X^I_{+} \ , \ \ \ 
\hat D_\mu X^I = \p_\mu X^I - X_+^I  B_\mu \ .
\ee
The Lagrangians $L_1, \ L_2$ are invariant under the non-compact gauge symmetry transformations
\be
\delta B_\mu =\p_\mu \Lambda \ ,\qquad \delta X^I = X_+^I  \Lambda  \ ,\qquad \delta X_-^I = \Lambda X ^I\ ,\qquad \delta X_+^I=0 \ .
\label{simme}
\ee
Note that $\delta \big(\tilde D_\mu X^I  \big)=0$ and that $\delta(\p_\mu X^I_{-} - X^IB_\mu)= \Lambda \hat D_\mu X^I$ while 
$\delta \big( \hat D_\mu X^I\big) =\Lambda\p_\mu X^I_{+}$. 
Therefore, the last terms  of eq.(\ref{xotto}) are also gauge invariant since they can be written as
\bea
(\p_\mu X^I_{-}-X^I B_\mu)\p^\mu X^I_{+}
- {X_{+}\cdot X\ov X^2_{+}}\hat D_\mu X^I\p^\mu X^I_{+} + {1\over 2}  ({X_{+}\cdot X\ov X^2_{+}})^2\p_\mu X^I_{+}\p^\mu X^I_{+}=  \nonumber\\
=(\p_\mu X^I_{-}-X^I B_\mu)\p^\mu X^I_{+} - {1\ov 2}\hat D_\mu X^I\hat D^\mu X^I +{1\ov 2}\tilde D_\mu X^I\tilde D^\mu X^I\ ,
\label{abid}
\eea
i.e. they are given by the same gauge-invariant combination appearing in the low energy lagrangian (\ref{lag1}) plus the
gauge-invariant term $\tilde D X\tilde DX$.
The full expression (\ref{abid}) vanishes for constant $X^I_{+}$.

The basic difference between the two non-linear Lagrangians $L_1$ and $L_2$ is that in the second case the kinetic term
$\p_\mu X_+^I \p^\mu X_-^I$ is outside the square root. The remaining terms have to be added to preserve gauge invariance and to preserve the connection with (\ref{lag1}) at low energies.
As we will see, in the non-abelian case, only the second Lagrangian $L_2$ can be constructed, because $X_+^I,X_-^I$ are $SU(N)$ singlets
and cannot be put inside the trace in a way preserving both $SU(N)$ and $B_\mu$ gauge invariance.

Following the method of \cite{mukhi}, we assume that $X_+^I$ takes an expectation value, so that $X_+^I$ is equal to constant vector $v^I $ plus a small fluctuation.
Then the Lagrangians $L_1$ and $L_2$  become
\be
L\equiv L_1 =L_2=  - T \sqrt{ -\det \big(\eta_{\mu\nu} + {1\ov T} (\p_\mu X^I -v^I  B_\mu ) (\p_\nu X^I-v^I  B_\nu )  \big) } + {1\over 2}\epsilon^{\mu\nu\rho}   B_\mu F_{\nu\rho} \ ,
\label{nove}
\ee
where we ignore terms with fluctuations which are suppressed at large $v^I$.

We can use the global $SO(8)$ symmetry to fix $v^I = v\delta_{I8}$. We get
\be
L=  - T \sqrt{ -\det \big(\eta_{\mu\nu} +{1\ov T} \p_\mu X^i  \p_\nu X^i + {1\ov T} (\p_\mu X^8 -v  B_\mu ) (\p_\nu X^8-v  B_\nu )  \big) } 
+ {1\over 2}\epsilon^{\mu\nu\rho}   B_\mu F_{\nu\rho} 
\label{dieci}
\ee
By choosing the gauge $X^8 = 0$ for the symmetry (\ref{simme}), and taking $v=g_{\rm YM}$, we finally obtain
\be
L= - T \sqrt{ -\det \big(\eta_{\mu\nu} + {1\ov T} \p_\mu X^i  \p_\nu X^i + {1\ov T} g^2_{\rm YM}  B_\mu    B_\nu  \big) } + {1\over 2}\epsilon^{\mu\nu\rho}   B_\mu F_{\nu\rho} \ .
\label{diecis}
\ee
This is precisely the previous Lagrangian (\ref{sei}). 

\section{Born-Infeld Lagrangian for Non-Abelian BF membrane }

Our starting point is  the  Lagrangian (\ref{myersL}) describing the dynamics of  $N$ coinciding D2 branes. 
Writing as before $\Phi^i=g_{\rm YM} X^i$, the D2 brane  Lagrangian is:
\be
L=-{1\over \lambda^2 g_{\rm YM}^2}{\rm STr} \sqrt{-\det (\eta_{\mu\nu}+\lambda^2 g_{\rm YM}^2 D_\mu X^i Q^{-1}_{ij}D_\nu X^j+\lambda F_{\mu\nu}) \det Q}\ ,
\label{D2}
\ee
where 
\be
Q^{ij}=\delta^{ij}+i\lambda g_{\rm YM}^2 [X^i,X^j]\ ,\qquad i,j=1,\cdots ,7\ .
\ee
Here we will make a simplifying assumption by considering only the symmetric part of $Q^{-1}_{ij}$, i.e. we write 
\be
{\rm STr}\sqrt{\cdots D_\mu X^i Q^{-1}_{ij}D_\nu X^j \cdots}\to {\rm STr}\sqrt{\cdots D_\mu X^i {Q^{-1}_{ij}+Q^{-1}_{ji}\ov 2}D_\nu X^j \cdots}
\ee
Due to the symmetrized trace prescription, by this assumption we only miss terms involving contractions of $D_\mu X^i Q^{-1}_{ij}D_\nu X^j$  and $F_{\mu\nu}$.\footnote{The  Lagrangian
that incorporates also the antisymmetric part of $D_\mu X^i Q^{-1}_{ij}D_\nu X^j$ was recently completed in \cite{Garousi}, after this paper appeared, 
following the same construction presented here.}

Therefore, by defining $g_{\mu\nu}\equiv\eta_{\mu\nu}+D_\mu X^i (Q^{-1})_{(ij)}D_\nu X^j$, where $(ij)$ denotes symmetrization,  we have that, inside the STr prescription,
$g_{\mu\nu}=g_{\nu\mu}$ and we can treat $g_{\mu\nu}$ as a metric.\footnote{Note
that $(Q^{-1})_{(ij)}$ is different from $(Q_{(ij)})^{-1}=\delta_{ij}$.}

\vskip0.5cm

We begin by showing that the D2 brane Lagrangian (\ref{D2}) has the equivalent form
\be
{\cal L} =
-T \ {\rm STr}  \sqrt{-\det \Big( \eta_{\mu\nu}+{1\ov T} D_\mu X^i  \big(Q^{-1} \big)_{(ij)} D_\nu X^j+{1\ov T}v^2{B_\mu B_\nu\ov\det Q} \Big)\det  Q}
+\Tr \Big({1\over 2 }\epsilon^{\mu\nu\rho} B_\mu F_{\nu\rho} \Big)\ ,
\label{bbbb}
\ee
with
\be
v=g_{\rm YM}\ .
\ee
First, we use the relation (\ref{cinque}) for $3\times 3$ matrices,
with $g_{\mu\nu}= \eta_{\mu\nu}+{1\ov T} D_\mu X^i  (Q^{-1})_{(ij)} D_\nu X^j$, and $K_\mu = v B_\mu/\sqrt{T\det Q} $,
and write, introducing a Lagrange multiplier $u$,
\bea
-T\sqrt{-\det (\eta_{\mu\nu}+{1\ov T} D_\mu X^i  (Q^{-1})_{(ij)} D_\nu X^j+{1\ov T}v^2{B_\mu B_\nu\ov\det Q})\det \ Q}
\nonumber\\
={1\ov 2u}T^2\det Q\det g+{1\ov 2u}\det g~Tv^2 B_\mu B_\nu g^{\mu\nu} 
-{u\ov 2}\ .
\label{multiplier}
\eea
Every term in the above expression is a (gauge-group) matrix. In the following manipulations we treat them as c-numbers, 
assuming that it is justified by the STr prescription. 

The equation of motion for $B_\mu$ gives
\be
g^{\mu\nu}B_\nu=-{u\ov 2}{\epsilon^{\mu\nu\rho} F_{\nu\rho}\ov Tv^2~\det g} \ .
\label{eqB}
\ee
Substituting back we get
\bea
-T\sqrt{-\det (\eta_{\mu\nu}+{1\ov T} D_\mu X^i  \big( Q^{-1} \big) _{(ij)} D_\nu X^j+{1\ov T}v^2{B_\mu B_\nu\ov\det Q})\det \ Q}
+{1\over 2 }\epsilon^{\mu\nu\rho} B_\mu F_{\nu\rho}=  \nonumber \\
={1\ov 2u}T^2\det Q\det g-{u\ov 2}(1+{|F|^2\ov 2 T v^2})\ ,
\label{BtoF}
\eea
where $|F|^2\equiv g^{\mu\mu'}g^{\nu\nu'}F_{\mu\nu}F_{\mu'\nu'}$ and we have made use of 
$g_{\mu\mu'}\epsilon^{\mu\nu\rho} F_{\nu\rho}\epsilon^{\mu'\nu'\rho'} F_{\nu'\rho'}=2\det g |F|^2$.
\vskip0.2cm

Solving for $u$, setting $v=g_{\rm YM}$ and using eqs. (\ref{tenz}), (\ref{tre}),
we finally obtain the D2 brane Lagrangian (\ref{D2}).

\medskip

Just as in the abelian case, the above Lagrangian (\ref{bbbb}) originates
from an M2 brane Lagrangian, where the term $B_\mu B_\nu$ arises from a term $D_\mu X^8 D_\nu X^8$.
The $SO(8)$ invariant starting point must be of the form
$\tilde D_\mu X^I  \tilde Q^{-1}_{IJ} \tilde D_\nu X^J$, $I,J=1,\cdots ,8$, where $\tilde Q_{IJ}$ and the covariant derivative
$\tilde D_\mu $ are to be determined.
The connection with the D2 brane Lagrangian (\ref{bbbb}) requires that, upon setting $X_+^I=v\delta_{I8}$, with $v=g_{\rm YM}$,
one gets
\be
\tilde D_\mu X^I  \tilde Q^{-1}_{IJ} \tilde D_\nu X^J 
 \to    D_\mu X^i Q^{-1}_{ij} D_\nu X^j + v^2{B_\mu B_\nu\ov\det Q}\ .
\ee
Therefore, 
\be
X_+^I=v\delta_{I8}\to \ \ \tilde Q^{ij}=Q^{ij}\ ,\qquad  \tilde Q^{i8}=\tilde Q^{8j}=0, \ \ \ \tilde Q^{88}=\det Q \ .  
\label{aqa}
\ee
Hence
\be
\det\tilde Q=(\det Q)^2\ .
\ee
One could in principle relax the condition $\tilde Q^{ij}=Q^{ij}$
in (\ref{aqa}) and impose the weaker condition 
$(\tilde Q^{-1})_{(ij)}=(Q^{-1})_{(ij)}$. However, it turns out that the simplest ansatz for $\tilde Q^{IJ}$ naturally gives $\tilde Q^{ij}=Q^{ij}$.

Invariance under the non-abelian $B_\mu $ gauge transformations (\ref{Bgauge}) 
is achieved by defining, just like in the abelian case,
\be
\tilde D_\mu X^I = \hat D_\mu X^I -{X_{+}\cdot X \over X^2_{+}}\p_\mu X^I_{+} \ ,
\label{bgau}
\ee
where $ \hat D_\mu X^I = D_\mu X^I-X^I_{+} B_\mu$  is the covariant derivative (\ref{covdev}) appearing in the low energy lagrangian
(now $D_\mu X^I=\p_\mu X^I+i [A_\mu ,X^i]$). 
It follows that $\delta \big(\tilde D_\mu X^I \big)=0$  under (\ref{Bgauge}). 
Recall that $X^I_{\pm}$ are $SU(N)$ singlets.

Let us now return to the general form of $\tilde Q^{IJ}$. This must be given in terms of $X_+^I$ and $X^I$ in a combination
invariant under the $B_\mu $-gauge transformations (\ref{Bgauge}).
It should not depend on $X_-^I$ in order to maintain the important property of the low energy BF membrane Lagrangian (\ref{lag1}) that interactions do not involve $X_-^I$
(this ensures, in particular, that $X_+^I,X_-^I$ do not propagate in loops \cite{GMR}). 
Some simple gauge-invariant $SO(8)$ tensors are $\delta^{IJ}$, $X_+^I X_+^J$, $X^K_{+}M^{IJK}$, where $M^{IJK}$ was defined in eq. (\ref{mijk}).
More general gauge-invariant operators involving $X_+^I$ and  $X^J$'s can be constructed by forming products 
$O_n\equiv X_+^{[J_1}X^{J_2}...X^{J_n]}$, where $[...]$ denotes complete antisymmetrization in all indices\footnote{This observation is due to M. Van Raamsdonk.}.
Then one can  define $SO(8)$ tensors $P^{IJ}_n=(O_n\cdot O_{n-2})^{IJ}$ or $R^{IJ}_n=(O_n\cdot O_{n})^{IJ}$ (in a short-hand notation, meaning that
all indices are contracted except two indices $I,J$).
The simplest gauge-invariant $SO(8)$ tensor $ \tilde Q^{IJ}$ satisfying the ``boundary" conditions (\ref{aqa}) is 
in fact of the form\footnote{The condition $\tilde Q^{ij}=Q^{ij}$ seems to leave (\ref{tilQ}) as the unique solution, 
since $X^K_{+}M^{IJK} =R_2^{JI}-R_2^{IJ}$, with $R_2^{IJ}=O_2^{IK}O_2^{KJ}$, is the only gauge-invariant operator which is antisymmetric in $IJ$ and quadratic in $X^I$. On the other hand, we have not found any simpler $\tilde Q^{IJ}$ from the weaker condition 
$(\tilde Q^{-1})_{(ij)}=(Q^{-1})_{(ij)}$.} 
\be
\tilde Q^{IJ}= a(X,X_{+})\ \delta^{IJ} +b(X,X_{+})\ X_+^I X_+^J +c(X,X_{+})\ {X^K_{+}M^{IJK}}\ ,
\label{tilQ}
\ee
where $a,b,c$ are gauge-invariant (and $SO(8)$ invariant) functions of $X^I,X^I_{+}$.

Imposing the condition (\ref{aqa})
for $ X_+^I=v\delta_{I8}$,   with $ v=g_{\rm YM}$ (noting that  $T^{-1/2}v=\lambda g_{\rm YM}^2$ and $M^{8ij}=v[X^i,X^j]$), then $\tilde Q^{IJ}$ is uniquely determined:
\be
\tilde Q^{IJ}\equiv  S^{IJ}+{X^I_{+}X^J_{+}\ov X^2_{+}}(\det (S) - 1 )
=(\delta^{IJ}-{X^I_{+}X^J_{+}\ov X^2_{+}}+{i \over \sqrt{T}} {X^K_{+}M^{IJK}\ov \sqrt{X^2_{+}}} )+{X^I_{+}X^J_{+}\ov X^2_{+}}\det (S)\ ,
\label{tQ}
\ee
where
\be
 S^{IJ}\equiv \delta^{IJ}+ {i \over \sqrt{T}}  {X^K_{+}M^{IJK}\ov \sqrt{X^2_{+}}} \ ,\qquad X^2_{+}=X^I_{+}X^I_{+}\ .
\ee
In the above formulas, it is understood that $\delta^{IJ}$ and $X_+^IX_+^J$ are multiplied by the identity matrix ${\cal I}_{N\times N}$. 

One can check that $~\tilde Q^{IJ}~$ is indeed invariant under $B_\mu$-gauge transformations (\ref{Bgauge}). 
Note that the expression (\ref{tQ})  involves a decomposition in a first term orthogonal to $X_+^I$ (since $X^I_{+}M^{IJK}X^K_{+}=0$ by virtue of the fact that  $M^{IJK}$ is completely antisymmetric), and a second term proportional to $X_+^I X_+^J$ (hence $X_+^I X_+^J \tilde Q^{IJ}=X_+^2 \det (S )$).

\medskip

One can check that 
\be
{X^L_{+}M^{LJK}\ov \sqrt{X^2_{+}}}{X^I_{+}M^{IKJ}\ov \sqrt{X^2_{+}}}=-{1\ov 3}M^{IJK}M^{IJK}
\ee
and
\bea
\Tr \bigg( {i \over \sqrt{T}}  {X^I_{+}M^{IJK}\ov \sqrt{X^2_{+}}} +{X^J_{+}X^K_{+}\ov X^2_{+}}\big( \det ( S) - 1 \big)\bigg)^n=
\Tr \bigg( {i \over \sqrt{T}} {X^I_{+}M^{IJK}\ov \sqrt{X^2_{+}}} \bigg)^n + \big( \det (S) - 1 \big)^n \\ \nonumber
\rightarrow
\Tr \log(\tilde Q^{IJ})=\Tr \log( S^{IJ})+ \log\big( \det (S) \big) \ \
\rightarrow
\det\tilde Q=\big(\det (S)\big)^2\ .
\eea

\vskip0.2cm

Thus we are led to the following  nonlinear Lagrangian for multiple M2 branes:
\bea
&&{\cal L} =
 - T\ {\rm STr} \bigg( \sqrt{ -\det \left(\eta_{\mu\nu} + {1\ov T} \tilde D_\mu X^I  \tilde Q^{-1}_{IJ}\tilde D_\nu X^J
\right) }(\det \tilde Q)^{1/4} \bigg)
+\Tr \big( {1\over 2}\epsilon^{\mu\nu\rho}   B_\mu F_{\nu\rho} \big)
\non\\
&&+ (\p_\mu X^I_{-}-\Tr(X^I B_\mu))\p^\mu X^I_{+}
-  \Tr \left( {X_{+}.X\ov X^2_{+}}\hat D_\mu X^I\p^\mu X^I_{+} - {1\over 2}  ({X_{+}. X\ov X^2_{+}})^2\p_\mu X^I_{+}\p^\mu X^I_{+}\right)
\label{M2inv}                                                         
\eea
The connection with the D2 brane Lagrangian is thus as follows. 
For $X_{+}^I=v \delta_{I8}$ we get $S^{ij}=Q^{ij}$, $S^{8i}=S^{i8}=0$, $S^{88}=1$, hence 
$\det S=\det Q$, $\det \tilde Q=(\det Q)^2$ and $(\tilde Q^{-1})_{ij}= Q^{-1}_{ij}$, $(\tilde Q^{-1})_{88}=1/\det Q$.
Then, by choosing the gauge $X^8=0$ we recover (\ref{bbbb}), which, by the steps (\ref{multiplier}), (\ref{eqB}), (\ref{BtoF}), can be connected
to the D2 brane Lagrangian (\ref{D2}).

As in the abelian case, the last term is 
added in order to match the low-energy Lagrangian. Note that it vanishes for constant $X_+^I$. Its origin is the 
non-abelian version of the gauge-invariant combination eq.(\ref{abid}).\footnote{
The appearance of  factors $X_+^2=X_+^IX_+^I$ in the Lagrangian (\ref{M2inv}), and the fact that the Yang-Mills coupling 
is $g_{\rm YM}^2= \langle X_+^IX_+^I \rangle $, may suggest an interpretation of $X_+^2$ as a radial coordinate representing
the center of mass position of the M2 branes \cite{Sen}.
However, this does not seem to be the precise role of $X_+^I$ in the Lagrangian (\ref{M2inv}). }

At the linearized approximation
\be
\det \tilde Q = \big( \det (S) \big)^2 \cong 1+ {1\ov T}{X^L_{+}M^{LJK}\ov \sqrt{X^2_{+}}} {X^I_{+}M^{IKJ}\ov \sqrt{X^2_{+}}}
= 1-{1\ov 3T}M^{IJK}M^{IJK}\ .
\label{lak}
\ee
Note that the factors $\sqrt{X^2_{+}}$ appearing in the denominator have canceled out. It can be easily shown that
this is the case to all orders, viz. all terms in the expansion of the potential $V=   T \ {\rm STr} \sqrt{\det (S)} $ in powers of $T^{-1}$
only contain non-negative powers of $X_+^2$.


\smallskip

Using (\ref{lak}), the Lagrangian (\ref{M2inv}) becomes, 
\bea
{\cal L}&=& - N T+ \Tr\big[ {1\over 2 }\epsilon^{\mu\nu\rho} B_\mu F_{\nu\rho}-{1\over 2}\hat D_\mu X^I\hat D^\mu X^I  +{1\ov 12}M^{IJK}M^{IJK}\big] 
+ (\p_\mu X^I_{-}-\Tr[X^I B_\mu ])\p^\mu X^I_{+}
\non\\
&+& O(l_p^3)
\eea
that is, we get the Lagrangian (\ref{lag1}).

\section{Fuzzy Funnel for M2-M5 brane Intersection}

In this section we compare a BPS solution of the low energy Lagrangian (\ref{lag1}) with an exact solution of
the  non-linear system (\ref{M2inv}). The solution generalizes the fuzzy funnel solution of \cite{constable1} describing $N$ D1 branes ending in a
D3 brane to eleven dimensions.
Studies of BPS solutions in the Bagger-Lambert system can be found in  \cite{Basu,berman,BL2,Lee,macca,Jeon,Passerini}.

\subsection{BPS solution in BF membrane model }

The BPS equations corresponding to the system (\ref{lag1}) are given by \cite{GMR,ver1,Mat}
\bea
\delta \Psi_+ &=& \p_\mu X_+^I \Gamma^\mu \Gamma^I \ \vareps =0\ ,
\non\\
\delta \Psi_- &=& \big( \p_\mu X_-^I - \Tr[B_{\mu } X^I ]\big) \Gamma^\mu \Gamma^I \vareps - {1\over 3}\Tr \big[X ^I 
X^J X^K \big]\Gamma^{IJK}\ \ \vareps =0\ ,
\label{bpsGMR}
\\
\delta \Psi  &=& \big( \p_\mu X^I- B_{\mu } X_+^I+ [ A_{\mu } ,X^I ] \big)
 \Gamma^\mu \Gamma^I \eps - X^I X^J X_+^K \Gamma^{IJK} \ \vareps =0\ .
\non
\eea
The  world-volume directions are $\s_{\hat 0},\ \s_{\hat 1},\ \s_{\hat 2}$ and they are identified with $0,9,10$ (so that
$ \Gamma^{\hat 0}=\Gamma^0,\ \Gamma^{\hat 1}=\Gamma^9,\ \Gamma^{\hat 2}=\Gamma^{10}$).
Here  $\vareps $ is an eleven-dimensional Majorana spinor satisfing the condition  $\Gamma_{\hat 0\hat 1\hat 2}\vareps =\vareps $.

To solve the first equation, we set $X_+^I=v\delta_{I8}$.
We then look for solutions with $B_{\mu}= A_{\mu}=0$ and set $X=X^I_a T^a$, $\Psi =\Psi_a T^a$, with $\Tr[T_a T_b ]= K \delta_{ab}$.
The remaining equations reduce to 
\bea
\delta \Psi_- &=& \p_\mu X_-^I \Gamma^\mu \Gamma^I \vareps -  {1\over 6}K\, C^{bcd} X_b^I 
X_c^J X_d^K \Gamma^{IJK}\ \ \vareps =0\ ,
\non\\
\delta \Psi_ a  &=&  \p_\mu X_a^I   \Gamma^\mu \Gamma^I \vareps -
{v\over 2} C_{\ \ \ a}^{bc} X_b^I  X_c^J   \Gamma^{IJ8} \ \vareps =0\ .
\label{bps2}
\eea
The system admits a solution with $SU(2)$ symmetry.  We set $T^i=\alpha^i,\ i=1,2,3$ to be   $SU(2)$ generators in some  $N\times N$ representation, so that
$C^{ijk}=2\eps^{ijk}$.  We then consider the ansatz
\be
X_a^I= f(\s )\delta_{aI}, \qquad a,I=1,...,3\ ,\qquad X_-^I= p(\s ) \delta_{I8}\ ,
\ee
where $\s\equiv \s_{\hat 1} $. This gives the equations
\be
p'(\s )= \mp 2 K f(\s )^3\ ,\qquad f'(\s )=\pm 2v f(\s )^2\ ,\
\label{solug}
\ee
and the conditions on the spinor
\be
\Gamma^{12389}\vareps = \mp \vareps\ .
\ee
The equation $f'(\s )=\pm 2v f(\s )^2$ is exactly the same equation that arises for the fuzzy funnel in the D1-D3 brane system 
(taking into account the normalization (\ref{nomal})).
The solution is given by
\be
v f(\s ) = \mp {1\over 2(\s_1-\s_{\infty })} \ ,
\ee
where $\s_\infty $ is an integration constant representing the position of the D3 brane.
Integrating the equation for $p$, we get
\be
p(\s )= \mp {K\over 8v^3(\s_1-\s_{\infty })^2} \ .
\ee
For an irreducible $N\times N$ $SU(2)$ representation $K = {1\over 3}N (N^2-1)$. 

\subsection{Funnel in non-linear M2 brane theory}

Here we discuss the funnel solution starting from  the
non-linear $M2$ brane Lagrangian (\ref{M2inv}).
The ansatz is: 
\bea
&& X^i=f(\sigma) \a^i ~, ~~ \ i=1,2,3\ , \ \ ~~X^I=0~~{\rm for}~~I>3 \ ,
\\ \nonumber
&& X^I_{+}=v(\sigma)\delta_{I8}~ , \ \ ~~X^I_{-}=p(\sigma)\delta_{I8} \\ \nonumber
&& B_\mu=0~,\ \ ~~F_{\mu\nu}=0\ ,
\eea
where, as before,  $\a^i$ are the $SU(2)$ generators in some  $N\times N$ representation, and $\sigma \equiv \sigma_{\hat 1}$ is a world-volume space coordinate.

With this ansatz the Lagrangian (\ref{M2inv}) becomes:
\be
{\cal L} = -T \ {\rm STr} \bigg(\sqrt{({\cal I}+{1\ov T} f'^2 \a^i Q^{-1}_{ij}\a^j)\det Q}\bigg)+ p'v'
\ee
where ${\cal I}_{N\times N}$ is the identity matrix,
and $Q^{ij}={\cal I}\delta^{ij} +{i\over \sqrt{T}} v^2f^2[\a^i,\a^j]$.

In the large $N$ limit, assuming the symmetrized trace prescription, we obtain
\be
 \a^i Q^{-1}_{ij}\a^j= C_2 {\cal I}\ ,\ \  \ \ \ \det Q={\cal I}+4T^{-1} f^4 v^2 C_2{\cal I} \ ,
\ee
where $C_2$ is the quadratic Casimir of the $SU(2)$ $N\times N$ representation. Therefore
\be
{\cal L} = -T N\sqrt{(1+{1\ov T} f'^2 C_2)(1+4 {1\ov T} f^4 v^2 C_2  )} + p'v' \ .
\ee
The variation with respect to $p$ gives $v=const$.
The variation with respect to $v$ gives the equation:
\be
p''+ 4vf^4 NC_2 \sqrt{1+{1\ov T} f'^2 C_2\ov 1+4 {1\ov T} f^4 v^2 C_2}=0\ .
\label{ppz}
\ee
One can substitute the second-order equation for $f$ by the condition 
\be
f' {\delta{\cal L}\ov\delta f'}+ p' {\delta{\cal L}\ov\delta p'}+v' {\delta{\cal L}\ov\delta v'}-{\cal L}=const
\rightarrow \sqrt{1+{1\ov T} f'^2 C_2\ov 1+4 {1\ov T} f^4 v^2 C_2}=const\ .
\ee
The last equation is solved by the solution of the first order equation:
\be
f'= \pm 2f^2v\ ,
\label{zuno}
\ee
whereby it follows that the equation (\ref{ppz}) for $p$ is equivalent to
\be
p'= \mp 2N {C_2\ov 3}f^3 = \mp 2K f^3 .
\label{zdos}
\ee
The above system of two first order equations (\ref{zuno}), (\ref{zdos}) is the same as eq. (\ref{solug}),  obtained by looking for a 
supersymmetric solution of the linearized Lagrangian.  Thus the BPS
solution of the leading order theory (\ref{lag1}) is also a solution of the full nonabelian M2 brane non-linear Lagrangian (\ref{M2inv}).

\section{Discussion}

Summarizing, we found the following non-linear Lagrangian
\bea
&&{\cal L} =
 - T\ {\rm STr} \bigg( \sqrt{ -\det \left(\eta_{\mu\nu} + {1\ov T} \tilde D_\mu X^I  \tilde Q^{-1}_{IJ}\tilde D_\nu X^J
\right) }(\det \tilde Q)^{1/4} \bigg)
+\Tr \big( {1\over 2}\epsilon^{\mu\nu\rho}   B_\mu F_{\nu\rho} \big)
\non\\
&&+ (\p_\mu X^I_{-}-\Tr(X^I B_\mu))\p^\mu X^I_{+} 
-  \Tr \left( 
{X_{+}\cdot X\ov X^2_{+}}\hat D_\mu X^I\p^\mu X^I_{+} - {1\over 2}  \Big( {X_{+}\cdot X\ov X^2_{+}} \Big)^2\p_\mu X^I_{+}\p^\mu X^I_{+}\right)
\label{Mcon}                                                         
\eea
where
\be
\tilde D_\mu X^I = \hat D_\mu X^I -{X_{+}\cdot X \over X^2_{+}}\p_\mu X^I_{+} \ ,
\label{bcon}
\ee
\be
\hat D_\mu X^I \equiv D_\mu X^I-X^I_{+} B_\mu , \ \  \ \  D_\mu X^I\equiv\p_\mu X^I+i[A_\mu, X^I]\ .
\label{covcon}
\ee
and $\tilde Q^{IJ}$ is defined in eqs. (\ref{tQ}), (\ref{mijk}).
It is invariant under $SU(N)$ gauge transformations and under the non-compact $B_\mu $ gauge transformations (\ref{Bgauge}).
The equation of motion for $X_-^I$ gives
\be
\p_\mu \p^\nu X_+^I=0\ .
\ee

The second line of eq. (\ref{Mcon}) --~which is gauge invariant by itself and vanishes for constant $X_+^I$~-- ensures the  match with the low-energy theory.
As pointed out  in section 2, the  kinetic term for $X_+^I,X_-^I$ appearing in the low-energy Lagrangian (\ref{lag1}) cannot be put inside the square root 
because $X_+^I,\ X_-^I$ are $SU(N)$  singlets. 
Since $\hat D_\mu X^I \hat D_\nu X^J$ alone is not gauge invariant  under $B_\mu $ gauge transformations (\ref{Bgauge}),
one is led to introduce the covariant derivative $\tilde D_\mu X^I$ to render the square-root term invariant. 
On the other hand, the factor $(\det \tilde Q)^{1/4}$  ensures that, after setting $X^I_{+}=v \delta_{I8}$,  the correct D2 Lagrangian (\ref{D2}) is reproduced, 
neglecting the antisymmetric part of $D_\mu X^i Q^{-1}_{ij}D_\nu X^j$  and modulo terms involving fluctuations of
$X^8_+$ which are suppressed at large $v$. These fluctuation terms are totally absent if 
 the shift symmetry $X_-^I\to X_-^I+c^I$ is gauged as in \cite{schwarz,Gomis2,mukhi2}
by  adding the term $-C^I_\mu\p^\mu X^I_{+}$.
Indeed, the equation  of motion of $C^I_\mu$ is $\p_\mu X^I_{+}=0$, which  sets $X^I_+$ to a constant value $v^I$.

\bigskip

In conclusion,  the Lagrangian (\ref{Mcon})  satisfies the following  properties:

\begin{itemize} 

\item $SO(8)$ invariance.

\item Invariance under the local gauge symmetries of the BF theory with algebra (\ref{bfalg}) (i.e. $SU(N)$ gauge invariance and $B_\mu$-gauge transformations (\ref{Bgauge})).

\item It contains just one dimensionful parameter $l_p^3$ (or $T=1/(4\pi^2 l_p^3)$), which disappears in the low energy approximation.

\item At low energies the Lagrangian (\ref{Mcon}) reduces to the bosonic part of the BF membrane Lagrangian (\ref{lag1}).

\item When $X_+^I$ takes a large expectation value the Lagrangian (\ref{Mcon})  gets connected to the  non-abelian D2 brane Lagrangian (\ref{myersL}).\footnote{See footnote 2.}

\item The supersymmetric fuzzy funnel is a solution of the non-linear Lagrangian (\ref{Mcon}) to all orders.
It does not receive any correction, just as it is the case for the D brane fuzzy funnel system describing the intersection of a D1 and a D3 brane \cite{constable1}.

\end{itemize}

\section*{Acknowledgments}

R.I. acknowledges the hospitality at the Department ECM of the
University of Barcelona where this work was initiated
This work is also supported by the European
EC-RTN network MRTN-CT-2004-005104. J.R. also acknowledges support by 
MCYT FPA 2007-66665 and CIRIT GC 2005SGR-00564.

\vfill\eject\null


\begin{thebibliography}{99}

\bibitem{BL1} 
  J.~Bagger and N.~Lambert,
  ``Gauge Symmetry and Supersymmetry of Multiple M2-Branes,''
  Phys.\ Rev.\  D {\bf 77}, 065008 (2008)
  [arXiv:0711.0955 [hep-th]].


\bibitem{Basu}
  A.~Basu and J.~A.~Harvey,
  ``The M2-M5 brane system and a generalized Nahm's equation,''
  Nucl.\ Phys.\  B {\bf 713}, 136 (2005)
  [arXiv:hep-th/0412310].


\bibitem{gus} 
  A.~Gustavsson,
  ``Selfdual strings and loop space Nahm equations,''
  JHEP {\bf 0804}, 083 (2008)
  [arXiv:0802.3456 [hep-th]].
  

\bibitem{papa}  G.~Papadopoulos,
  ``M2-branes, 3-Lie Algebras and Plucker relations,''
  JHEP {\bf 0805}, 054 (2008)
  [arXiv:0804.2662 [hep-th]].


\bibitem{gaun}  J.~P.~Gauntlett and J.~B.~Gutowski,
  ``Constraining Maximally Supersymmetric Membrane Actions,''
  arXiv:0804.3078 [hep-th].

\bibitem{Tong} N.~Lambert and D.~Tong,
  ``Membranes on an Orbifold,''
  arXiv:0804.1114 [hep-th].
  
\bibitem{Distler}
  J.~Distler, S.~Mukhi, C.~Papageorgakis and M.~Van Raamsdonk,
  ``M2-branes on M-folds,''
  JHEP {\bf 0805}, 038 (2008)
  [arXiv:0804.1256 [hep-th]].
  
\bibitem{GMR} 
J.~Gomis, G.~Milanesi and J.~G.~Russo,
 ``Bagger-Lambert Theory for General Lie Algebras,''
  JHEP {\bf 0806}, 075 (2008)
  [arXiv:0805.1012 [hep-th]].


\bibitem{ver1}  S.~Benvenuti, D.~Rodriguez-Gomez, E.~Tonni and H.~Verlinde,
  ``N=8 superconformal gauge theories and M2 branes,''
  arXiv:0805.1087 [hep-th].

\bibitem{Mat}  P.~M.~Ho, Y.~Imamura and Y.~Matsuo,
  ``M2 to D2 revisited,''
  JHEP {\bf 0807}, 003 (2008)
  [arXiv:0805.1202 [hep-th]].
  
\bibitem{mukhi} 
S.~Mukhi and C.~Papageorgakis,
  ``M2 to D2,''
  JHEP {\bf 0805}, 085 (2008)
  [arXiv:0803.3218 [hep-th]].

\bibitem{cecotti} S.~Cecotti and A.~Sen,
  ``Coulomb Branch of the Lorentzian Three Algebra Theory,''
  arXiv:0806.1990 [hep-th].

\bibitem{Fig}  P.~de Medeiros, J.~M.~Figueroa-O'Farrill and E.~Mendez-Escobar,
  ``Metric Lie 3-algebras in Bagger-Lambert theory,''
  arXiv:0806.3242 [hep-th].

\bibitem{ASS} 
 M.~Ali-Akbari, M.~M.~Sheikh-Jabbari and J.~Simon,
  ``Relaxed Three-Algebras: Their Matrix Representations and Implications for
  Multi M2-brane Theory,''
  arXiv:0807.1570 [hep-th].

\bibitem{ver3} H.~Verlinde,
  ``D2 or M2? A Note on Membrane Scattering,''
  arXiv:0807.2121 [hep-th].
%
 
\bibitem{schwarz} M.~A.~Bandres, A.~E.~Lipstein and J.~H.~Schwarz,
  ``Ghost-Free Superconformal Action for Multiple M2-Branes,''
  arXiv:0806.0054 [hep-th].


\bibitem{Gomis2} J.~Gomis, D.~Rodriguez-Gomez, M.~Van Raamsdonk and H.~Verlinde,
  ``Supersymmetric Yang-Mills Theory From Lorentzian Three-Algebras,''
  arXiv:0806.0738 [hep-th].

\bibitem{mukhi2}  B.~Ezhuthachan, S.~Mukhi and C.~Papageorgakis,
  ``D2 to D2,''
  JHEP {\bf 0807}, 041 (2008)
  [arXiv:0806.1639 [hep-th]].



\bibitem{tseytlin}
 A.~A.~Tseytlin,
  ``On non-abelian generalisation of the Born-Infeld action in string
  theory,''
  Nucl.\ Phys.\  B {\bf 501}, 41 (1997)
  [arXiv:hep-th/9701125].
  

\bibitem{myers}
  R.~C.~Myers,
  ``Dielectric-branes,''
  JHEP {\bf 9912}, 022 (1999)
  [arXiv:hep-th/9910053].


\bibitem{BST}
E.~Bergshoeff, E.~Sezgin and P.~K.~Townsend,
  ``Supermembranes and eleven-dimensional supergravity,''
  Phys.\ Lett.\  B {\bf 189}, 75 (1987).



\bibitem{abjm}  O.~Aharony, O.~Bergman, D.~L.~Jafferis and J.~Maldacena,
  ``N=6 superconformal Chern-Simons-matter theories, M2-branes and their
  gravity duals,''
  arXiv:0806.1218 [hep-th].


\bibitem{klebanov}  
M.~Benna, I.~Klebanov, T.~Klose and M.~Smedback,
  ``Superconformal Chern-Simons Theories and AdS4/CFT3 Correspondence,''
  arXiv:0806.1519 [hep-th].

\bibitem{schwarz2} M.~A.~Bandres, A.~E.~Lipstein and J.~H.~Schwarz,
  ``Studies of the ABJM Theory in a Formulation with Manifest SU(4)
  R-Symmetry,''
  arXiv:0807.0880 [hep-th].



\bibitem{Xie}
  T.~Li, Y.~Liu and D.~Xie,
  ``Multiple D2-Brane Action from M2-Branes,''
  arXiv:0807.1183 [hep-th].


\bibitem{kluson}
J.~Kluson,
  ``D2 to M2 Procedure for D2-Brane DBI Effective Action,''
  arXiv:0807.4054 [hep-th].


\bibitem{mat1} 
 P.~M.~Ho and Y.~Matsuo,
  ``M5 from M2,''
  JHEP {\bf 0806}, 105 (2008)
  [arXiv:0804.3629 [hep-th]].


\bibitem{mat2}  P.~M.~Ho, Y.~Imamura, Y.~Matsuo and S.~Shiba,
  ``M5-brane in three-form flux and multiple M2-branes,''
  JHEP {\bf 0808}, 014 (2008)
  [arXiv:0805.2898 [hep-th]].


\bibitem{BT1}
 I.~A.~Bandos and P.~K.~Townsend,
  ``Light-cone M5 and multiple M2-branes,''
  arXiv:0806.4777 [hep-th].

\bibitem{BT2} 
 I.~A.~Bandos and P.~K.~Townsend,
  ``SDiff Gauge Theory and the M2 Condensate,''
  arXiv:0808.1583 [hep-th].



\bibitem{constable1}
  N.~R.~Constable, R.~C.~Myers and O.~Tafjord,
  ``The noncommutative bion core,''
  Phys.\ Rev.\  D {\bf 61}, 106009 (2000)
  [arXiv:hep-th/9911136].


\bibitem{town}  E.~Bergshoeff and P.~K.~Townsend,
  ``Super D-branes,''
  Nucl.\ Phys.\  B {\bf 490}, 145 (1997)
  [arXiv:hep-th/9611173].

\bibitem{nicolai} H.~Nicolai and H.~Samtleben,
  ``Chern-Simons vs. Yang-Mills gaugings in three dimensions,''
  Nucl.\ Phys.\  B {\bf 668}, 167 (2003)
  [arXiv:hep-th/0303213];
B.~de Wit, H.~Nicolai and H.~Samtleben,
  ``Gauged supergravities in three dimensions: A panoramic overview,''
  arXiv:hep-th/0403014.



\bibitem{Sen}
S.~Banerjee and A.~Sen,
  ``Interpreting the M2-brane Action,''
  arXiv:0805.3930 [hep-th].


\bibitem{berman}
  D.~S.~Berman and N.~B.~Copland,
  ``Five-brane calibrations and fuzzy funnels,''
  Nucl.\ Phys.\  B {\bf 723}, 117 (2005)
  [arXiv:hep-th/0504044].


\bibitem{BL2}
 J.~Bagger and N.~Lambert,
  ``Comments On Multiple M2-branes,''
  JHEP {\bf 0802}, 105 (2008)
  [arXiv:0712.3738 [hep-th]].



\bibitem{Lee}
  K.~Hosomichi, K.~M.~Lee and S.~Lee,
  ``Mass-Deformed Bagger-Lambert Theory and its BPS Objects,''
  arXiv:0804.2519 [hep-th].

 
\bibitem{macca}
C.~Krishnan and C.~Maccaferri,
  ``Membranes on Calibrations,''
  JHEP {\bf 0807}, 005 (2008)
  [arXiv:0805.3125 [hep-th]].



\bibitem{Jeon}
  I.~Jeon, J.~Kim, N.~Kim, S.~W.~Kim and J.~H.~Park,
  ``Classification of the BPS states in Bagger-Lambert Theory,''
  JHEP {\bf 0807}, 056 (2008)
  [arXiv:0805.3236 [hep-th]].

\bibitem{Passerini}
  F.~Passerini,
  ``M2-Brane Superalgebra from Bagger-Lambert Theory,''
  arXiv:0806.0363 [hep-th].

\bibitem{Garousi}
 M.~R.~Garousi,
  ``On non-linear action of multiple M2-branes,''
  arXiv:0809.0985 [hep-th].

  
 \end{thebibliography}
\end{document}